\documentclass[fleqn,10pt]{wlscirep}
\usepackage[utf8]{inputenc}
\usepackage[T1]{fontenc}

\usepackage{bm}

\title{Mind the gap in university rankings: a complex network approach towards fairness}

\author[1,2]{Loredana Bellantuono}
\author[2]{Alfonso Monaco}
\author[3,2,*]{Nicola Amoroso}
\author[4,$\S$]{Vincenzo Aquaro}
\author[5,6,$\P$]{Marco Bardoscia}
\author[7,8]{Annamaria Demarinis Loiotile}
\author[8,2]{Angela Lombardi}
\author[9,2]{\\ Sabina Tangaro}
\author[8,2]{Roberto Bellotti}

\affil[1]{Dipartimento di Scienze Mediche di Base, Neuroscienze e Organi di Senso, Università degli Studi di Bari Aldo Moro, 70124, Bari, Italy}
\affil[2]{Istituto Nazionale di Fisica Nucleare, Sezione di Bari, 70125, Bari, Italy}
\affil[3]{Dipartimento di Farmacia-Scienze del Farmaco, Università degli Studi di Bari Aldo Moro, 70125, Bari, Italy}
\affil[4]{Division for Public Institutions and Digital Government, United Nations Department of Economic and Social Affairs (DESA), New York, NY, 10017, USA}
\affil[5]{Bank of England, London EC2R 8AH, United Kingdom}
\affil[6]{Department of Computer Science, University College London, London WC1E 6BT, United Kingdom}
\affil[7]{Dipartimento di Ingegneria Elettrica e dell'Informazione, Politecnico di Bari, 70125, Bari, Italy}
\affil[8]{Dipartimento Interateneo di Fisica, Università degli Studi di Bari Aldo Moro, 70126, Bari, Italy}
\affil[9]{Dipartimento di Scienze del Suolo, della Pianta e degli Alimenti, Università degli Studi di Bari Aldo Moro, 70126, Bari, Italy}

\affil[*]{nicola.amoroso@uniba.it}

\affil[$\S$]{The designations employed and the presentation of the material in this paper do not imply the expression of any opinion whatsoever on the part of the United Nations concerning the legal status of any country, territory, city or area, or of its authorities, or concerning the delimitation of its frontiers or boundaries. The designations ``developed'' and ``developing'' economics are intended for statistical convenience and do not necessarily imply a judgment about the state reached by a particular country or area in the development process. The term ``country'' as used in the text of this publication also refers, as appropriate, to territories or areas. The views expressed are those of the individual authors of the paper and do not imply any expression of opinion on the part of the United Nations.}

\affil[$\P$]{Any views expressed are solely those of the author(s) and so cannot be taken to represent those of the Bank of England or to state Bank of England policy.}

\begin{abstract}
University rankings are increasingly adopted for academic comparison and success quantification, even to establish performance-based criteria for funding assignment. However, rankings are not neutral tools, and their use frequently overlooks disparities in the starting conditions of institutions. In this research, we detect and measure structural biases that affect in inhomogeneous ways the ranking outcomes of universities from diversified territorial and educational contexts. Moreover, we develop a fairer rating system based on a fully data-driven debiasing strategy that returns an equity-oriented redefinition of the achieved scores. The key idea consists in partitioning universities in similarity groups, determined from multifaceted data using complex network analysis, and referring the performance of each institution to an expectation based on its peers. Significant evidence of territorial biases emerges for official rankings concerning both the OECD and Italian university systems, hence debiasing provides relevant insights suggesting the design of fairer strategies for performance-based funding allocations.
\end{abstract}
\begin{document}

\flushbottom
\maketitle
%
%
\thispagestyle{empty}


\section*{Introduction}

Far beyond being mere educational institutions, universities play an outstanding role in citizen formation, scientific and technological progress, elaboration of new models of economy and society, which makes them drivers of development \cite{1}. Universities contribute to the society both on an individual scale, as they often represent one of the most effective social elevators, and on a collective scale, as they build knowledge to face global challenges, with a specific attention, increasing in recent years, to the ones related to sustainable development goals \cite{2}. The rise in relevance of rankings concerning academic institutions is a rather recent phenomenon, that emerged since the late 1980s \cite{3}, mainly due to the demand for information on academic quality by potential students, triggered by the worldwide expansion of the access to higher education \cite{4}. However, rankings have become an increasingly internalized tool for comparison and success quantification far beyond the matter of student’s choice, influencing researchers, employers and, most relevantly, academic evaluators and companies.

University rankings are compiled taking into account the variety of missions higher education is called to, which are not limited to teaching and research, but also involve the so called ``third mission'' \cite{8} or ``knowledge transfer'' \cite{9,10}, namely the impact of the academic activity on the community in which the institution is embedded. Efficient higher education institutions can actually prompt a spillover process in which regions collect knowledge and human resources, that can contribute to foster their economic progress \cite{7,11,12,15}. Therefore, it is evident that the evaluation of academic activity should be configured as a multi-purpose assessment \cite{13}, which takes into account not only the scientific impact, but also the benefits brought to the territory, measured by an empirical quantification of the third mission outcomes \cite{14}.

Rankings, whatever they are meant to measure, are not neutral tools, and their use has a series of relevant drawbacks. Problematic aspects mainly stem from the effects of positive feedback between the prestige of an institution, certified by its ranked score, and the possibility to receive public funding on an awarding basis or to attract investments by private companies \cite{13,16,17}. The strong causal relation between ranking outcomes and funding triggers off undesired phenomena. The first problem is reactivity to rankings, namely the development of adaptation strategies to gain competitive advantages with respect to the evaluation criteria, leading to academic conformism \cite{13,18,19,20,21,22}. Another critical aspect is represented by territorial biases, that reward universities placed in an advantageous socioeconomic context \cite{7,23,24,25,26,27}, that can be, for example, more receptive than others with respect to third mission activities. The third issue is the onset of a ``Matthew effect'', that, through the feedback between ranking and funding, consolidates existing gaps in third mission \cite{28}, internationalization \cite{29,30}, research \cite{5,6}, scholarships \cite{31}, and even diffusion of scientific ideas \cite{32}. 

Lately, rankings have prompted deep changes in the higher education system, affecting resource distribution, decision making and status definition \cite{3,18,33,34}. However, they fail to capture individual specificities and tend to marginalize parts of the academic community whose distinctive traits are not suited to the general rating framework \cite{31,35}. Most importantly, the indicators used to compile university rankings can be affected by biases related to the specific institution mission and to territorial features, such as the presence of human resources, stakeholders, and a vital industrial background. Though the methodology behind them is criticized, and their overall role is questioned \cite{36}, rankings nowadays represent a consolidated evaluation framework, mainly due to their simplicity and practicality, combined with the lack of suitable alternatives \cite{37}. On one hand, students reasonably require ``absolute'' ratings of universities, to aim for the best affordable education and to maximize the possibilities of a good employment in a wealthy environment after graduation. On the other hand, a different kind of rankings, in which the effects of structural factors is mitigated, can be relevant for academic evaluators and policy makers. Such redefined rankings could actually help identify both virtuous cases of outstanding institutions emerging in a difficult context, and cases in which the performance is below expectations, which therefore require intervention. Therefore, it is urgent to define transparent, data-driven, shared and reproducible procedures to evaluate academic performance, taking into account taking into account the effect of structural features, such as the territorial embedding of universities and their educational mission. 

The objectives of our research are: 1) analyzing university rankings in order to detect biases, determined by either the territorial context or the educational offer; 2) quantifying the effect of the aforementioned biases on the performance of each university; 3) defining a new score in which biases are mitigated. We remark that the expressions ``unbiased'', ``debiased/debiasing'', ``fairer" used in the following will be exclusively referred to the kind of biases considered in our work, namely those related to territorial features and educational offer. To achieve these results, we will follow a vision that stems from the concept of economic complexity \cite{38,39,40,41}, according to which the performance of a country in a given economic task is the result of several underlying context variables, also including intangible assets. In order to generalize the idea to our problem, we will model the academic environment as a complex system, in which interconnections among its elements prompt the emergence of similarities and properties in an unsupervised and data-driven way. Complex network theory provides the mathematical machinery to formalize such a description \cite{42,43,44,45}, and represents a multidisciplinary tool, increasingly used to investigate real-world systems consisting of nontrivially interconnected constituents. Application fields include economics \cite{41,46,47,48,49,50}, human mobility \cite{51,52}, neuroscience \cite{53,54,55,56}, genetics \cite{57,58}, just to mention a few. Moreover, recent studies are endowing the complex network toolbox with new instruments, such as multilayer networks \cite{59} and network potentials \cite{60,61}. Complex network methods have already been applied to the problem of ranking analysis, to formalize the detection of competitiveness patterns among sport teams \cite{62} and universities \cite{63}, and to encompass multifaceted context-based information in the evaluation of performances achieved by countries \cite{64}.

In this work, we construct network models to analyze two case studies: the set of universities from member countries of Organization for Economic Co-operation and Development (OECD), appearing in the 2021 Times Higher Education (THE) rankings \cite{65}, and the Italian tertiary education system, surveyed through the rankings compiled by CENSIS (Centro Studi Investimenti Sociali) for the academic year 2019/2020 \cite{66}. The motivation of studying the Italian university environment as a national case is rooted in the geographical polarization of the country between wealthy regions in the northern part and struggling ones in the south. In particular, more than one fourth of Italian population lives in one of the four ``convergence regions'', whose GDP per capita is lower than 75\% of the European Union average. On the other hand, since university evaluation criteria are uniform throughout the nation, it is reasonable to expect some sort of bias in rankings. 

In both cases of OECD and Italy, we construct two complex networks, based on the territorial similarity and the educational offer similarity, respectively. We first determine whether there is a tendency for institutions belonging to similar territories, or providing analogous educational offers, to achieve comparable scores in a given ranking. Since our findings reveal the presence of a bias in many specific rankings, we construct a suitable unbiased ``reference system'', determined in an unsupervised way, to evaluate a university's performance. The rationale of this approach is the detection of those institutions that share common features, and can therefore be fairly compared. Finally, we introduce a quantifier of the bias that affects each single university, and define for each index a new fair ranking, in which bias has been removed. We present the details of our findings in the Result section, and comment their implications in the Discussion. The Materials and Methods section contains technical details on data collection, network construction, and implementation of the complex-network methods.

\section*{Results}

\begin{figure}
\centering
\includegraphics[width=0.9\linewidth]{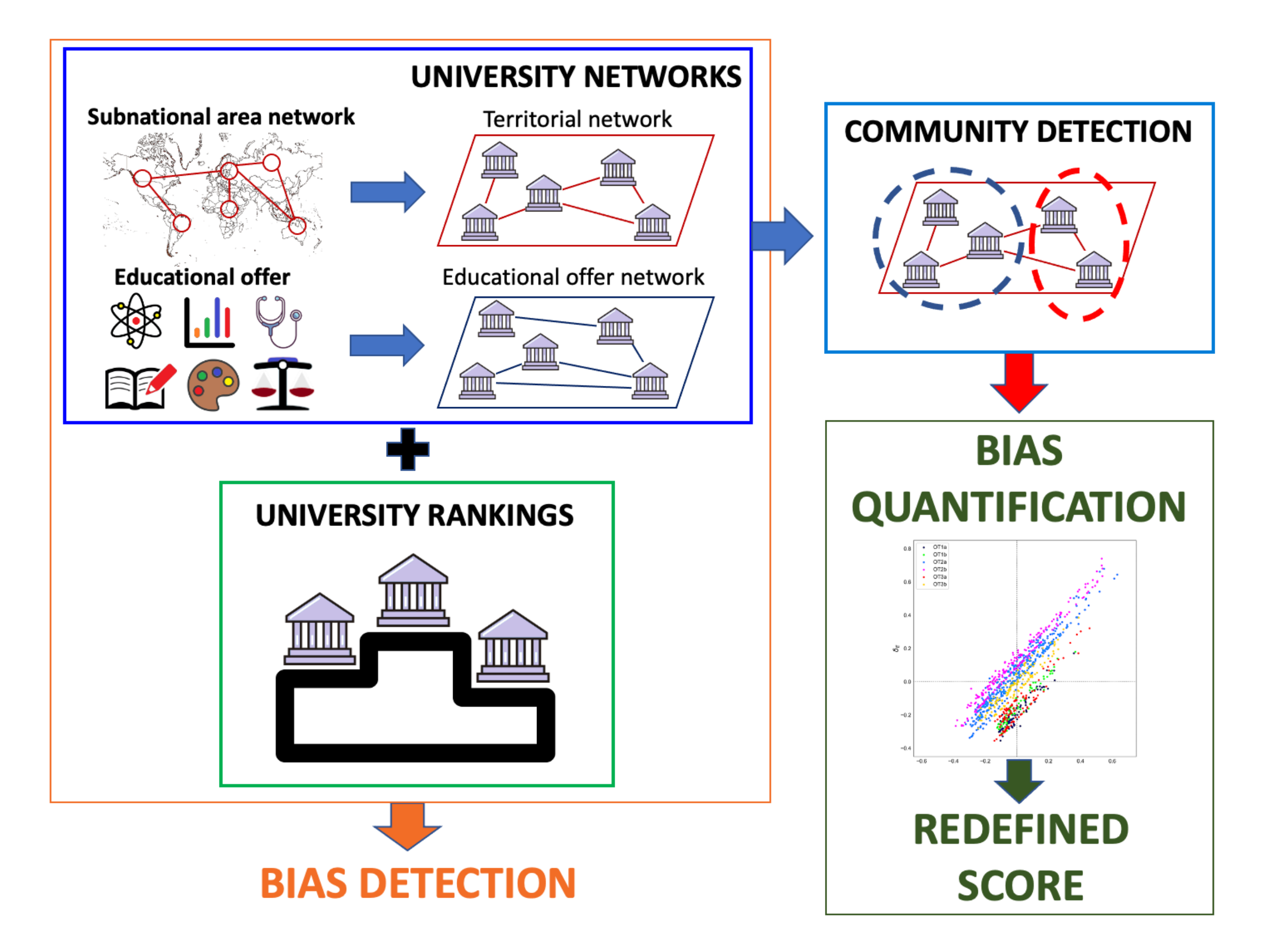}
\caption{Scheme representing the workflow of the analysis. Indicators on the territorial features (used to construct a network of subnational areas) and information on the educational offer of academic institutions constitute the basis to build a territorial network and an educational offer network of universities, respectively. Network analysis enables to detect and quantify biases in university rankings, providing the necessary information to define a new score in view of a fair evaluation.}
\label{fig:schema}
\end{figure}

The goals of this work are detecting structural biases in university rankings, measuring their impact on scores and defining a debiasing strategy to develop a fairer rating system. A scheme of the workflow implemented in the present research is displayed in Fig. 1. We model the university environment as a pair of complex networks, in which connections are determined by similarity of territorial features and educational offers. Evidence of biases emerges from this framework for a variety of official rankings. To quantify the effects of these biases on the outcomes achieved by single universities, we partition the academic ecosystem in homogeneous groups through community detection algorithms, and perform comparisons among peers therein. This approach allows to define fairer scores, in which the detected bias is decoupled from performance evaluation. In this section, we describe the main findings of our analysis, separating the presentation into two case studies: the former concerning the best-ranked universities of OECD countries, and the latter focusing on the whole ecosystem of Italian universities.

\subsection*{OECD university networks}

We consider 1088 universities, ranked in the THE website \cite{65} and distributed in 343 OECD subregions at a subnational scale (see Materials and Methods for details). The THE ranking measures the performance of universities on a global scale by providing an overall rating that combines different scores related to the fundamental aspects of the university mission: teaching, research, citations, industry income and international outlook. To analyze the performance achieved by the institutions in the light of their starting conditions, we set up two different networks: (i) a territorial network determined by the features of the region in which each institution is based, thus encoding information that is external to the academic system; (ii) an educational offer network, that incorporates internal features of the academic environment.

\subsubsection*{Territorial OECD university network} 

To highlight similarities and differences between territories, the most natural choice is to build a \textit{subnational area network}, whose nodes coincide with the aforementioned 343 OECD subregions. Affinity among territories is determined from 97 indicators on geography, economy and society, and modeled by means of weighted edges. A pair of subregions is connected by an edge provided the Pearson correlation between the sets of their related territorial indicators is statistically significant; the procedure, described in detail in Materials and Methods, returns 29415 edges, weighted by the correlation itself. Starting from the OECD subnational area network, we construct the proper \textit{territorial university network}, consisting of 1088 nodes representing universities, and 351186 edges between them, related to the similarity between the corresponding subregions. We assign edges and weights of the territorial network according to this criterion: universities belonging to two regions related by a weighted edge in the subnational area network are connected by an edge of the same weight; different universities belonging to the same region are connected by an edge of unitary weight. In this way, we embed the university network in the territorial context.

\subsubsection*{Educational offer OECD university network} 

The second international university network is constructed from the educational offer of the same 1088 universities that appear in the THE rankings. The 539305 edges connecting each pair of institutions are weighted by a measure of the overlap between their didactical offers. Since the presence of a given sector in the educational offer range represents a binary variable, we use the Dice index to quantify such an overlap. The implementation of this procedure is described in detail in Materials and Methods.

\subsubsection*{Bias detection in THE rankings through assortativity analysis}

We begin the analysis of OECD university networks by investigating whether the similarity among THE scores is related to the similarity, in terms of either territory or educational offer, among the institutions that achieve them. The tendency in a network to connect nodes with similar attribute values is quantified by \textit{assortativity} (see Materials and Methods section for definition and characterization). As reported in the first row of Table~\ref{tab:THE_assortativity}, the territorial network is assortative with respect to the \textit{THE overall} score. This finding represents strong evidence of an evaluation bias related to the geographical context: universities based in similar territories tend to achieve comparable scores. Furthermore, the territorial network is assortative in a statistically significant way with respect to all the specific THE ranking dimensions, with the largest values registered for \textit{THE citations} and \textit{THE international outlook}. In the case of the educational offer network, instead, even the few significant values are negligible. Accordingly, this finding suggest that educational specificities do not play a relevant role in determining a positive outcome of an academic institution in the ranking. A qualitative manifestation of the different tendencies discussed above can be observed in the scatter plots shown in Supplementary Figure~S1.

\begin{table}
\centering
\begin{tabular}{|l|r|r|}
\hline
\ & Territorial network & Educational offer network \\
\hline
THE overall score & $\bm{0.109 \pm 0.001}$ & $\bm{0.003 \pm 0.001}$ \\
\ & ($p<10^{-9}$) & ($p=0.007$) \\ 
\hline
THE teaching & $\bm{0.043 \pm 0.001}$ & $0.002 \pm 0.001$ \\
\ & ($p<10^{-9}$) & ($p=0.044$) \\ 
\hline
THE research & $\bm{0.059 \pm 0.001}$ & $0.002 \pm 0.001$ \\
\ & ($p<10^{-9}$) & ($p=0.015$) \\ 
\hline
THE citations & $\bm{0.143 \pm 0.001}$ & $\bm{0.004 \pm 0.001}$ \\
\ & ($p<10^{-9}$) & ($p=10^{-4}$) \\ 
\hline
THE industry income & $\bm{0.015 \pm 0.001}$ & $\bm{0.003 \pm 0.001}$ \\
\ & ($p<10^{-9}$) & ($p=0.003$) \\ 
\hline
THE international outlook & $\bm{0.147 \pm 0.001}$ & $0.002 \pm 0.001$ \\
\ & ($p<10^{-9}$) & ($p=0.038$) \\ 
\hline
\end{tabular}
\caption{\label{tab:THE_assortativity}Assortativity of the territorial network and the educational offer network of the OECD case study with respect to the following THE rankings: \textit{overall}, \textit{teaching}, \textit{research}, \textit{citations}, \textit{industry income}, \textit{international outlook}. For each assortativity value, the standard error and $p$-value, computed according to the Student $t$-distribution hypothesis, are provided (see Materials and Methods section for details); significant assortativity values ($p<10^{-2}$) are highlighted in boldface.}
\end{table}

Communities in OECD university networks. Community detection allows to collect universities that share homogeneous features in a network built from mutual similarity relations\cite{44,45}, providing the basis for a comparison among peers. Thus, it represents the most crucial step to go from bias detection to bias quantification. The application of hierarchical community detection (see Materials and Methods for implementation details) on the territorial network provides the following optimal partition:  
\begin{itemize}
    \item OT1a: 80 universities in Argentina, Brazil, Colombia, Costa Rica, Mexico, Peru;
    \item OT1b: 78 universities in Chile, Mexico, Turkey, United States (south-east);
    \item OT2a: 430 universities in Australia, Western Europe, Israel, Japan, New Zealand;
    \item OT2b: 241 universities in Canada, United States, Australia, New Zealand, Japan, Estonia, Spain (Balearic Islands);
    \item OT3a: 122 universities in Eastern Europe, Korea;
    \item OT3b: 137 universities in Southern and Central Europe, Korea. 
\end{itemize}
The geographical distribution of the above communities is reported in Supplementary Figure~S2, generated with the MapChart online tool \cite{mapchart}. As for the educational offer network, hierarchical community detection returns:
\begin{itemize}
    \item OE1a: 128 universities with predominant engineering and computer science areas, and underrepresented humanities, health and social science areas;
    \item OE1b: 173 universities with predominant science, engineering and economics areas;
    \item OE2a: 221 universities with a very generalized range of educational offer, not including veterinary science;
    \item OE2b: 206 universities with a very generalized range of educational offer, differing from OE2a due to the overrepresentation of “veterinary science” and “agriculture and forestry” areas, present in 100\% and 92\% of institutions, respectively;
    \item OE3a: 179 universities with a generalized range of educational offer, but underrepresented engineering areas;
    \item OE3b: 48 universities with an educational offer strongly focused on the health area;
    \item OE3c: 133 universities with predominant economics, humanities and social science areas.
\end{itemize}
It is worth noticing that the underrepresentation of engineering in the universities belonging to community OE3a is related in around two thirds of cases to the presence of a technical university (belonging to communities OE1a-OE1b) in the same territory. The community nomenclature, referring to both the territorial and educational offer networks, is explained in Materials and Methods. The full list of universities with their community membership is reported in Supplementary Data S1, while numerical details of hierarchical community detection and information on the intermediate-level partitions are provided in the Supplementary Section 1.2.

\subsection*{Quantifying biases and redefining scores in THE rankings} 

Community membership represents the main tool to quantify biases and fairly evaluate the performance of academic institutions in rankings. Once we have clustered universities in homogeneous OT and OE communities, we introduce new indicators that highlight the relationship between the success of an institution in rankings and either the territory where it is based or its educational offer.

\begin{figure}
    \centering
    \includegraphics[width=0.5\textwidth]{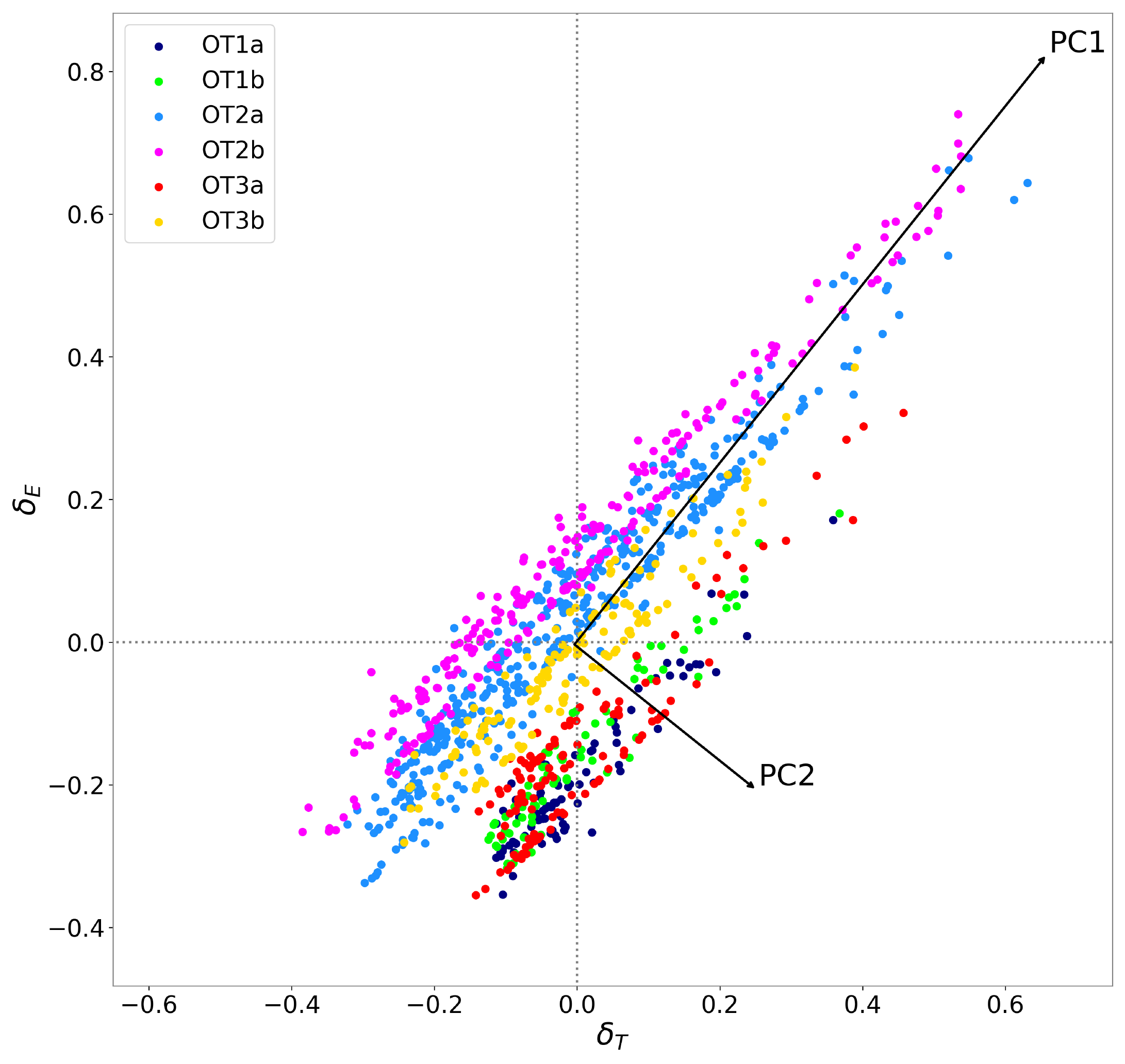}
    \caption{Territorial bias in \textit{THE overall} ranking. In this scatter plot each dot corresponds to an OECD university, and its coordinates along the horizontal and vertical axes represent the debiasing parameters. The values $\delta_T$ and $\delta_E$ assess the results achieved by the institution in the \textit{THE overall} ranking, respectively by comparison with the rest of the OT and OE community it belongs to. Each dot in the scatter plot is colored according to its OT community membership. The arrangement of different colors in the scatter plot indicates the presence of a territorial bias. The arrows indicate the direction of the principal components PC1 (positive slope) and PC2 (negative slope), with their length proportional to the corresponding standard deviations. }\label{fig:scatterTHEoverall}
\end{figure}

For a given ranked index $I$ we associate to each university $u$ two \textit{debiasing parameters} $\delta_T(u)$ and $\delta_E(u)$, respectively referred to the territorial network and the educational offer network. These parameters evaluate the performance of an institution by comparison with the rest of the community it belongs to. For each university $u$ belonging to the territorial community $C_T$ and to the educational offer community $C_E$, we associate debiasing parameters to a ranked index $I$ as follows
\begin{equation}\label{eq:delta}
    \delta_S(u) = I(u) - \frac{\sum_{v\in C_S} w_{uv}^S I(v)}{\sum_{v\in C_S} w_{uv}^S}, \qquad \text{with } S=T,E
\end{equation}
namely as the difference between the index value $I(u)$ of the considered institution and the average of the indexes $I(v)$ in the rest of the community, weighted by the $(u,v)$ edge weight $w_{uv}^S,\, (S=T,E)$ of the considered network. Notice that community peers that are weakly connected to $u$ give a negligible contribution to the average appearing in Eq.~\eqref{eq:delta}. A positive (negative) value of the debiasing parameter indicates a better (worse) performance than the average expectation for universities that are similar, in terms of either territory or educational offer. Debiasing parameters do not represent \textit{per se} a measure of bias. Their joint distribution, instead, can highlight systematic advantages for members of specific communities in a given ranked index. The scatterplot of $(\delta_T,\delta_E)$ values related to the \textit{THE overall} score in Figure~\ref{fig:scatterTHEoverall}, shows an evident grouping among territorial communities. An analogous arrangement is not present at all if points in the scatterplot are partitioned according to their educational offer community membership.

The distribution of $(\delta_T,\delta_E)$ values in Figure~\ref{fig:scatterTHEoverall}, as well as those obtained from the sectorial THE dimensions, reported in Supplementary Figure~S3, indicates that the grouping in terms of territorial communities mostly occurs along a direction that is orthogonal to the one of maximal variance. To quantify such a tendency, isolate the territorial bias affecting university rankings, and define a fairer rating, we determine the principal components of the distributions in the $(\delta_T,\delta_E)$ planes referred to all THE scores. Actually, as we will shortly demonstrate, the first principal component (PC1) represents a redefined ranking, in which geographical influence is mitigated, while the second one (PC2) provides a measure of the territorial dragging effect on the performance in the original ranking. To support this interpretation, we discuss in detail the case of THE overall. Here, the component PC2, accounting for $8.6$\% of the total variance, is markedly anticorrelated (Pearson correlation $-0.523$, with $p<10^{-9}$ from the exact distribution test) with the GDP per capita PPP (Gross Domestic Product per capita at Purchasing Power Parity) related to the OECD subregions where the main seats of the universities are located. This result indicates that the arrangement of territorial communities in Figure~\ref{fig:scatterTHEoverall} follows a wealth hierarchy: therefore, universities with higher values of PC2 are experiencing the most detrimental effect on their performance due to a disadvantaged territorial context. On the other hand, PC1, accounting for $91.4$\% of the total variance, shows a weaker positive correlation ($0.367$, with $p<10^{-9}$ from the exact distribution test) with the GDP per capita PPP. This outcome should also be compared with the Pearson correlation between the GDP and the original ranking ($0.481$, with $p<10^{-9}$), which is intermediate between the absolute values of the correlations with PC1 and PC2. This property is shared by all the sectorial THE dimensions, as shown in Supplementary Table~S1. Moreover, most importantly, the assortativity of the territorial university network with respect to PC1 ($0.054$) is significantly reduced compared to that of the original ranking, showing that PC1 represents a fairer rating system, in which the effect of the territorial bias is substantially mitigated. Actually, the parametrization of points in Figure~\ref{fig:scatterTHEoverall} in terms of the principal components leads to incorporating most of the bias in PC2: the assortativity of the territorial university network with respect to this direction ($0.227$) is about twice the value obtained for the original ranking. The discussed behavior is actually common to all THE dimensions, as shown in Supplementary Table~S1. Top ten performers in PC1 and PC2 related to THE overall are displayed in Tables~\ref{tab:THEoverall_PC1}-\ref{tab:THEoverall_PC2}, respectively, while the full lists, also including the principal component values with respect to each ranking dimension, are contained in Supplementary Data~S2.

The first quartile of largest improvements in PC1 with respect to the original ranking is composed of 97 universities from community OT3a, 70 from OT1a, 68 from OT1b, 26 from OT3b, 9 from OT2a, none from OT2b. From this result, we can observe how shifting from the original ranking to PC1 highlights the performance of universities that emerge in comparatively difficult development contexts. Therefore, one of the merits of the PC1 redefined ranking is the recognition of universities that, despite being in relatively disadvantaged territories, are able to achieve good results compared to their peers in both the territorial and the educational offer community.

\begin{table}
\centering
\begin{tabular}{|l|l|l|l|}
\hline
\textbf{University} & \textbf{Subregion} & \textbf{Country} & \textbf{PC1} \\
\hline
California Institute of Technology & California & United States & $0.916\,(+3)$ \\
\hline
University of Oxford & South East England & United Kingdom & $0.902\,(-1)$ \\
\hline
Massachusetts Institute of Technology & Massachusetts & United States & $0.884\,(+2)$ \\
\hline
Imperial College London & Greater London & United Kingdom & $0.877\,(+7)$ \\
\hline
Stanford University & California & United States & $0.873\,(-3)$ \\
\hline
University of Cambridge & East of England & United Kingdom & $0.871\,(0)$ \\
\hline
ETH Zurich  & Zurich & Switzerland & $0.847\,(+7)$ \\
\hline
Princeton University & New Jersey & United States & $0.837\,(+1)$ \\
\hline
Harvard University & Massachusetts & United States & $0.837\,(-6)$ \\
\hline
University of California, Berkeley & California & United States & $0.793\,(-3)$ \\
\hline
\end{tabular}
\caption{\label{tab:THEoverall_PC1}Top ten universities in the ranking of the principal component PC1 associated to the \textit{THE overall} score. In the last column, the number in round brackets indicates the position variation in the PC1 ranking with respect to the original one.}
\end{table}

\begin{table}
\centering
\begin{tabular}{|l|l|l|l|}
\hline
\textbf{University} & \textbf{Subregion} & \textbf{Country} & \textbf{PC2} \\
\hline
Lomonosov Moscow State University & Moscow Oblast & Russia & $0.195$ \\
\hline
Universidad Nacional del Litoral & Santa Fe & Argentina & $0.184$ \\ 
\hline
University of Campinas & Sao Paulo & Brazil & $0.181$ \\
\hline
University of Costa Rica & Central & Costa Rica & $0.179$ \\
\hline
University of Alabama at Birmingham & Alabama & United States & $0.175$ \\
\hline
University of S\~ao Paulo & Sao Paulo & Brazil & $0.174$ \\
\hline
Jagiellonian University & Lesser Poland & Poland & $0.168$ \\
\hline
Federal University of S\~ao Paulo (UNIFESP) & Sao Paulo & Brazil & $0.165$ \\
\hline
Hacettepe University & Ankara & Turkey & $0.163$ \\ 
\hline
Tomsk State University & Tomsk Oblast & Russia & $0.163$ \\
\hline
\end{tabular}
\caption{\label{tab:THEoverall_PC2}Top ten universities in the ranking of the principal component PC2 associated to the \textit{THE overall} score.}
\end{table}

\subsection*{Italian university networks}

The second part of the analysis is dedicated to the Italian university environment as a whole. The system consists of 92 institutions of diversified size and educational target. The Italian university rankings examined in our research are compiled by CENSIS\cite{64}, with the aim of providing a guideline to the choice of future students. Due to the target, these rankings evaluate aspects that have a relevant impact on the quality of students’ academic experience, such as the service provision, scholarship and structure availability, communication through digital instruments, internationalization, and employability of graduates. The results for each specific dimension are combined into a global score (\textit{CENSIS overall}). Following a similar procedure as in the OECD case, we will perform a deep investigation of the connection between the scores achieved by Italian academic institutions in CENSIS rankings and both the territorial features and the educational offer.

\subsubsection*{Territorial Italian university network} 

The analysis starts from a network of subnational areas, corresponding to the Italian provinces (\textit{province}), a statistical subdivision of the country of outstanding relevance, midway between regions and municipalities. The 53 provinces in which the Italian universities are distributed constitute the nodes of the subnational area network, in which 854 weighted edges are constructed according to a criterion of similarity, quantified by the Pearson correlation between 121 socio-economic indicators (see Materials and Methods for details). The proper territorial network, in which universities are encoded as nodes, is constructed in the same way as in the OECD case, and consists of 92 nodes and 2396 weighted edges.

\subsubsection*{Educational offer Italian university network} 

A second university network is constructed, as in the OECD case, based on the similarity among the educational offers provided by the different institutions (see Materials and Methods section for details). The variety of Italian higher education ranges from large general-purpose universities, to those small and with very circumscribed offer, passing through large but specialized ones such as polytechnics, focused on engineering, and private institutions mainly oriented to economics and law. The educational offer network is made of 92 nodes and 2007 weighted edges.

\subsubsection*{Bias detection in CENSIS rankings through assortativity analysis} 

We study the Italian academic system by means of an assortativity analysis which is analogous to the one implemented in the OECD case. The most relevant of the results reported in Table~\ref{tab:CENSIS_assortativity} is that the territorial network is assortative with respect to the \textit{CENSIS overall} score, indicating a statistically significant territorial bias, while the educational offer network's assortativity is negligible. Concerning the dimensions composing the \textit{CENSIS overall} ranking, the territorial bias is extremely relevant in the case of employability, an index that has no counterpart among the THE scores, and is high also in the case of international outlook, as in the corresponding THE ranking. As in the international case, no significant assortativity is found for any CENSIS ranking dimension in the educational offer network. The quantitative results on assortativity find a qualitative counterpart in the scatter plots shown in Supplementary Figure~S4.

\begin{table}
\centering
\begin{tabular}{|l|r|r|}
\hline
\ & Territorial network & Educational offer network \\
\hline
CENSIS overall score & $\bm{0.289 \pm 0.018}$ & $-0.019 \pm 0.017$ \\
\ & ($p<10^{-9}$) & ($p=0.275$) \\ 
\hline
CENSIS services & $0.027 \pm 0.019$ & $-0.002 \pm 0.017$ \\
\ & ($p=0.161$) & ($p=0.954$) \\ 
\hline
CENSIS scolarships & $-0.017 \pm 0.019$ & $-0.010 \pm 0.017$ \\
\ & ($p=0.388$) & ($p=0.541$) \\ 
\hline
CENSIS structures & $\bm{0.077 \pm 0.019}$ & $-0.020 \pm 0.017$ \\
\ & ($p=6\cdot 10^{-5}$) & ($p=0.243$) \\ 
\hline
CENSIS communication and & $\bm{0.122 \pm 0.019}$ & $-0.012 \pm 0.017$ \\
digital services & ($p<10^{-9}$) & ($p=0.475$) \\ 
\hline
CENSIS international outlook & $\bm{0.288 \pm 0.018}$ & $0.007 \pm 0.017$ \\
\ & ($p<10^{-9}$) & ($p=0.667$) \\ 
\hline
CENSIS employability & $\bm{0.436 \pm 0.022}$ & $-0.003 \pm 0.019$ \\
\ & ($p<10^{-9}$) & ($p=0.865$) \\ 
\hline
\end{tabular}
\caption{\label{tab:CENSIS_assortativity}Assortativity of the territorial network and the educational offer network of the Italian case study with respect to the following CENSIS rankings: \textit{overall}, \textit{services}, \textit{scholarships}, \textit{structures}, \textit{communication and digital services}, \textit{international outlook}, \textit{employability}. For each assortativity value, the standard error and $p$-value, computed according to the Student $t$-distribution hypothesis, are provided (see Materials and Methods section for details); significant assortativity values ($p<10^{-2}$) are highlighted in boldface.}
\end{table}

\subsubsection*{Communities in Italian university networks} 

Hierarchical community detection is performed for the Italian university networks, in the same spirit and with the same procedure as in the OECD case. The following partition of the territorial university network is obtained:
\begin{itemize}
    \item IT1a: 25 universities in center-north provinces with a small administrative center;
    \item IT1b: 35 universities in center-north provinces, mostly with a large or historically relevant administrative center;
    \item IT2a: 13 universities in center-south and Sardinia;
    \item IT2b: 19 universities in the south.
\end{itemize}
Such a partition reflects once more the long-standing social and economic gap between north and south of Italy (see Supplementary Figure~S5 \cite{mapchart}). As concerns the educational offer network, community detection provides
\begin{itemize}
    \item IE1: 31 small, telematic universities, oriented to law, economics or foreign languages;
    \item IE2: 44 medium-to-large general-purpose universities;
    \item IE3: 9 polytechnic and small engineering-oriented universities;
    \item IE4: 8 research hospitals and health-oriented small universities.
\end{itemize}
The full list of universities with their community membership in both the territorial and educational offer networks is reported in Supplementary Data~S3. The community nomenclature is explained in Materials and Methods. Numerical details of hierarchical community detection and the intermediate-level partitions that lead to the aforementioned results are reported, for both networks, in the Supplementary Section~2.2.

\begin{figure}
    \centering
    \includegraphics[width=0.5\textwidth]{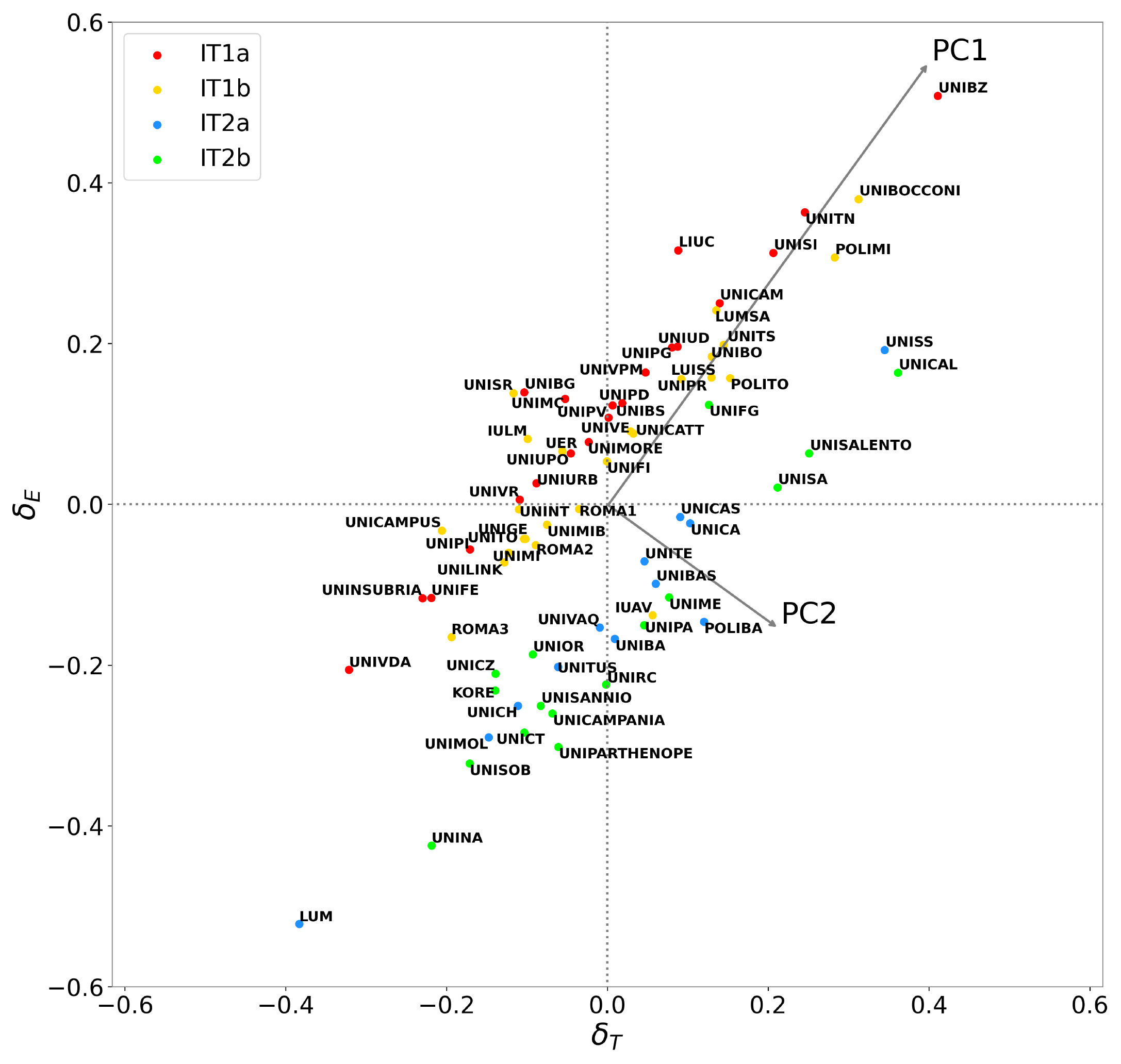}
    \caption{Territorial bias in \textit{CENSIS overall} ranking. In this scatter plot each dot corresponds to an Italian university, and its coordinates along the horizontal and vertical axes represent the debiasing parameters. The values $\delta_T$ and $\delta_E$ assess the results achieved by the institution in the \textit{CENSIS overall} ranking, respectively by comparison with the rest of the IT and IE community it belongs to. Each dot in the scatter plot is colored according to its IT community membership. The arrangement of dots and their color distribution in the scatter plot indicate the existence of two clusters with a very neat geographical characterization, separated by a gap that represents the territorial bias. The arrows indicate the direction of the principal components PC1 (positive slope) and PC2 (negative slope), with their length proportional to the corresponding standard deviations. }\label{fig:scatterCENSISoverall}
\end{figure}

\subsubsection*{Quantifying biases and redefining scores in CENSIS rankings} 

Even in the Italian case study, we adopt communities and the debiasing parameters as the main tool to implement a context-aware assessment of the scores achieved by universities in rankings. The scatter plot in Figure~\ref{fig:scatterCENSISoverall} represents universities in the $(\delta_T,\delta_E)$ plane in the case of the \textit{CENSIS overall} rating; analogous plots associated to the sectorial dimensions of CENSIS ranking are displayed in Supplementary Figure~S6. The arrangement of points in Figure~\ref{fig:scatterCENSISoverall} bears similarities to that shown in Fig~\ref{fig:scatterTHEoverall} for the OECD case study, but differs in a crucial point: in the $(\delta_T,\delta_E)$ plane referred to the overall CENSIS rating, two well-separated clusters are present, apparently distributed along parallel lines. These clusters are geographically characterized, in the sense of a north-south polarization. 

The principal components of point distributions in the $(\delta_T,\delta_E)$ plane provide a redefined ranking (PC1) and a measure of the territorial dragging effect (PC2). This interpretation is supported by results analogous to those obtained in the international case. In particular, for \textit{CENSIS overall}, the component PC2, accounting for $12.8$\% of the total variance, is strongly anticorrelated to the average per capita available income of the province hosting the main university seat (Pearson correlation $-0.636$, with $p=10^{-9}$ from the exact distribution test). On the other hand, PC1, accounting for $87.2$\% of the total variance, shows a positive and weaker correlation with the same wealth indicator ($0.405$, with $p=3\cdot 10^{-4}$ from the exact distribution test). Moreover, the assortativity value $0.113$ of the territorial network with respect to PC1 is strongly mitigated compared to the one related with the original ranking, while the assortative behavior concentrates on PC2 ($0.450$). Also in the Italian case, the Pearson correlation between the GDP and the original ranking ($0.553$, with $p=3\cdot 10^{-7}$), is intermediate between the absolute values of the correlations with PC1 and PC2. As reported in Supplementary Table~S2, the described phenomenology is found, with few exceptions, in the sectorial CENSIS ranking dimensions. Top ten performers in PC1 and PC2 related to \textit{CENSIS overall} are displayed in Tables~\ref{tab:CENSISoverall_PC1}-\ref{tab:CENSISoverall_PC2}, respectively, while the full lists, also including the principal component values with respect to each ranking dimension, are contained in Supplementary Data~S4. The top quartile of largest improvements from the original ranking to PC1 is composed of 7 universities from community IT2a, 7 from IT2b, 2 from IT1b, none from IT1a. Therefore, also in the Italian case, the redefined ranking highlights the performance of universities that emerge in comparatively difficult development contexts.

\begin{table}
\centering
\begin{tabular}{|l|l|l|}
\hline
\textbf{University} & \textbf{Province of main seat} & \textbf{PC1} \\
\hline
Free University of Bolzano & Bolzano & $0.652\,(0)$ \\
\hline
``Luigi Bocconi'' University of Milano & Milano & $0.491\,(+1)$ \\
\hline
University of Trento & Trento & $0.438\,(-1)$ \\
\hline
Milano Politecnico & Milano & $0.414\,(0)$ \\
\hline
University of Siena & Siena & $0.374\,(0)$ \\
\hline
University of Sassari & Sassari & $0.358\,(+5)$ \\
\hline
University of Calabria & Cosenza & $0.345\,(+7)$ \\
\hline
``Carlo Cattaneo University -- LIUC & Varese & $0.308\,(+4)$ \\
\hline
University of Camerino & Macerata & $0.285\,(-3)$ \\
\hline
``Maria SS.\ Assunta'' Free University -- LUMSA & Roma & $0.275\,(+5)$ \\
\hline
\end{tabular}
\caption{\label{tab:CENSISoverall_PC1}Top ten universities in the ranking of the principal component PC1 associated to the \textit{CENSIS overall} score. In the last column, the number in round brackets indicates the position variation in the PC1 ranking with respect to the original one.}
\end{table}

\begin{table}
\centering
\begin{tabular}{|l|l|l|}
\hline
\textbf{University} & \textbf{Province of main seat} & \textbf{PC2} \\
\hline
University of Calabria & Cosenza & $0.195$ \\
\hline
Bari Politecnico & Bari & $0.181$ \\ 
\hline
University of Sassari & Sassari & $0.165$ \\
\hline
University of Salento & Lecce & $0.165$ \\
\hline
University of Salerno & Salerno & $0.158$ \\
\hline
University of Messina & Messina & $0.129$ \\
\hline
``Mediterranea'' University of Reggio Calabria & Reggio di Calabria & $0.129$ \\
\hline
``Parthenope'' University of Napoli & Napoli & $0.126$ \\
\hline
IUAV University of Venice & Venezia & $0.125$ \\
\hline
University of Palermo & Palermo & $0.124$ \\ 
\hline
\end{tabular}
\caption{\label{tab:CENSISoverall_PC2}Top ten universities in the ranking of the principal component PC2 associated to the \textit{CENSIS overall} score.}
\end{table}

The gap between the two clusters appearing in Fig.~\ref{fig:scatterCENSISoverall}, measured along the PC2 direction, is numerically relevant: actually, it corresponds to $2.8$ times the average standard deviation of the PC2 distributions within each cluster. This result is obtained by quantifying the gap as the distance $0.147$ between the two peaks of the distribution of the principal component PC2, after checking its consistency with a bimodal Gaussian by means of a Gaussian Mixture model (see Materials and Methods for details). The idea of a systematic separation between clusters is further corroborated by independently checking that the respective distributions in the $(\delta_T,\delta_E)$ plane are fitted by two regression lines that are parallel within their standard errors:
\begin{equation}
\delta_E = (0.908 \pm 0.013) \delta_T + (0.097 \pm 0.003)
\end{equation}
for the upper cluster, and
\begin{equation}
\delta_E=(0.922 \pm 0.011) \delta_T + (-0.164 \pm 0.002)
\end{equation}
for the lower one, with the uncertainties determined by 10-fold cross validation. The average vertical distance $\Delta\delta_E$ between these lines can be interpreted as a measure of the north-south structural gap separating institutions that equally perform with respect to their own territory: given a pair of universities belonging to different clusters and characterized by the same $\delta_T$, regardless of its value, the university in the center-north cluster tends to have $\delta_E$ larger by $\Delta\delta_E = 0.260$ (which amounts to $25.2$\% of the total range of $\delta_E$) than that of the university in the center-south cluster.

\section*{Discussion}

In this work, we have achieved two relevant results: measuring the impact of the territory on the scores of universities in rankings, and decoupling this bias from the definition of performance, thus developing a fairer rating system. Nowadays, awareness of the effect of structural inequalities on performance evaluation is increasing, as testified by the many different attempts to detect and mitigate it. On one hand, the same rating agencies try to enact normalization strategies, that are only partially effective and require an independent testing of the residual bias. On the other hand, various studies have used regression methods to investigate the impact of possible underlying structural factors (such as GDP per capita, R{\&}D expenditure and English as a first language, to mention a few) on a university's score \cite{67,68,bornmann2013multilevel,69,daraio2015rankings}. This approach, recently employed even to perform bias removal \cite{70}, relies on the choice of a necessarily restricted and arbitrary set of aggregated indicators as possible bias sources. In our work, we overcome this limitation by following the paradigm of economic complexity, which allows to refer the performance of a university to a multifaceted representation of its context, determined by a large number of indicators. We also remark that our analysis is characterized by a higher geographical resolution than the state of the art; in particular, the case of OECD universities is investigated here on a subnational scale, whereas previous research on international rankings employed data available on a national basis.

We are able to determine the bias affecting the score of each single university (PC2), and a fairer ranking (PC1) in which the effect of structural inequalities is substantially mitigated. The quantities PC1 and PC2 are computed from the newly introduced debiasing parameters, which allow to refer the performance of each university to an expectation based on its peers, namely other universities located in a similar territorial context or providing an analogous educational offer. To the best of our knowledge, the proposed approach represents an unprecedented case of unsupervised and fully data-driven debiasing in rankings.

Universities achieving the largest placement improvements in PC1 with respect to the original overall rankings belong to comparatively disadvantaged territorial communities. Not surprisingly, the same geographical areas host universities with the highest values of PC2, namely those most negatively affected by the territorial bias. Top performers in PC1 are still based in the most developed communities, indicating that the excellence of some academic institutions persists even after bias removal. Nonetheless, a pair of European universities achieve significant upgrades among top-10 placements in PC1 related to \textit{THE overall}, and some universities from the south of Italy manage to bridge the gap and appear among top-10 PC1 associated to \textit{CENSIS overall}.

The analysis reveals that the territorial bias generally affects the score of OECD universities in all the THE rankings. This behavior is emphasized in the case of \textit{THE international outlook}: actually, infrastructures and cultural connections that characterize important and wealthy areas provide a much stronger boost to academic internationalization than those of peripheric and poorer zones \cite{68,71}. Territorial bias is relevant also in \textit{THE citations}, confirming a Matthew effect already known from previous literature \cite{70}: the reputational stock of eminent academic institutions, usually embedded in a wealthy context, attracts human and financial resources to foster new high-quality research, but at the same time induces biases in scientific production and paper citations, fueling a self-reinforcement mechanism. Instead, \textit{THE research} and \textit{THE industry income} are characterized by a comparatively low territorial bias: this rather counterintuitive result is due to a normalization with respect to GDP per capita PPP in the index definitions, that is capable of partially decoupling the bias from performance evaluation.

The case of Italian universities is characterized by the presence a gap in the distribution of PC2 related to the \textit{CENSIS overall} ranking and to some specific dimensions (see Supplementary Figure~S6), indicating a neat separation between positively and negatively biased universities. Such a feature has profound and structural reasons, related to the economic, historical and social gap between north and south of the country. The absence of a similar phenomenology in the international case indicates that OECD subregions are distributed according to a more smeared spectrum of development. The Italian north-south polarization becomes particularly striking in the case of \textit{CENSIS employability}: low employability is actually, with few exceptions, a widespread problem that graduates from universities based in south of Italy have to face, thus representing a crucial factor in determining internal student mobility. Finally, it is interesting to remark that \textit{CENSIS international outlook} is also affected by a strong territorial bias, as well as its THE counterpart. 

In this paper we have observed that the territorial context affects by a relevant amount an academic institution's performance, which, if positive, can contribute to further improvements through self-reinforcing awarding mechanisms. At this point, a natural question arises as to how much the advantageous features of a territory are determined by the presence of an outstanding university. This problem, that requires specific data and further analytic tools, represents an interesting research perspective that can be addressed in future works.

\section*{Materials and Methods}

\subsection*{Experimental design}

The goal of our analysis is to provide an interpretation to the score of universities in rankings, that takes into account both the territory in which they are embedded and their educational targets. Specifically, we aim at shedding light on the influence that is exerted on an academic institution by the context it belongs to. The analysis, replicated for the best-ranked institutions of OECD countries and for the Italian university environment, is made of the following steps:
\begin{itemize}
    \item collecting 1) values of territorial indicators, combining in a reasonable way data abundance and recentness, 2) information on the educational offer of universities, 3) higher-education rankings;
    \item constructing a pair of complex networks with nodes representing universities, the one based on territorial similarity and the other based on educational offer similarity, for both the international (OECD) and the Italian case;
    \item quantifying the assortativity of each university network with respect to the scores in higher-education rankings;
    \item using the unsupervised partition of these networks, obtained by community detection algorithms, to divide the university environment in homogeneous groups in terms of territory or educational offer, in order to compare the performance of each institution with those of its peers.
\end{itemize}
In the following, we detail the implementation of the aforementioned process, leaving more strictly technical aspects to the Supplementary Information.

\subsection*{Data collection and preprocessing}

Due to their heterogeneous nature, the data employed to construct the networks are collected from different sources. In particular, we use data from the following datasets:
\begin{itemize}
    \item OECD subregions data. Territorial indicators are collected from the OECD Regional Statistics database \cite{72,73}. We choose to associate each university seat to the region it belongs to at the Territorial Level 2 (TL2) of subdivision. For data availability reasons, we make exceptions to this rule, considering TL3 regions for Estonia and Latvia. The missing values of infant mortality for the regions of Brazil, and of life expectancy at birth for the regions of Brazil and Argentina, are collected from Global Data Lab (dataset version updated to November 2020), that focuses on emerging economies \cite{74}. The total number of international indicators is 103. For each indicator and subregion the most recent value available has been considered. The full list of OECD subregional indicators is reported in Supplementary Table~S3.
    \item Educational offer data of the best-ranked universities in the world, scraped from the THE website \cite{65}. The full list of the educational offer categories available at universities in the THE rankings is reported in Supplementary Table~S4.
    \item Two datasets, prepared by the Italian National Institute of Statistics (ISTAT), of territorial indicators on welfare \& sustainability (\textit{Indicatori benessere e sostenibilit\`a}) \cite{75}, and on development policies (\textit{Indicatori territoriali per le politiche di sviluppo}) \cite{76}, updated to November 2019. These datasets contain a total of 144 indicators referred to each Italian province, from which we exclude ``Average per capita income (EUR)''. For each indicator, data from the most recent year available are considered, with possible integrations of missing entries borrowed from at most 5 years before. The list of indicators along with their reference years is reported in Supplementary Table~S5.
    \item Data collected by the Italian Ministry of University and Research (MUR) in the survey on the educational offer of Italian universities, indicating which degree categories (\textit{classi di laurea}), determined by both the degree level and sector, are available at each institution \cite{77}. The classification of Italian degree categories according to the International Standard Classification of Education (ISCED) \cite{78} is reported in Supplementary Table~S6.
\end{itemize}
Indicators for the territorial analysis are subjected to a selection, which eliminates those correlated to another indicator by a Pearson correlation larger than $0.98$, as they carry redundant information. In cases of statistical redundancy, we retain the indicator with the wider data availability. This selection leaves 97 and 121 indicators for the territorial analysis of international and Italian universities, respectively. The selected indicators are finally normalized in order to compute Pearson correlation between different subregions and develop the complex network model. More precisely, before performing the linear rescaling of an indicator in the interval $[0,1]$, the values exceeding the 99th percentile from above and the 1st percentile from below are replaced by the reference percentiles, in order to mitigate the effect of outliers.

On the other hand, rankings are retrieved from the following sources:
\begin{itemize}
    \item the 2021 rankings of world universities compiled by THE \cite{65}; while the overall score is publicly provided only for the best 200 positions, the dimensions that contribute to it, namely teaching, research, citations, industry income and international outlook, are available for 1526 universities; the overall score is then computed following the instructions on the THE website; of all the ranked universities, we focus on 1088 institutions, located in OECD countries, for which a reasonable amount of territorial data is available;
    \item the academic year 2019/2020 rankings of Italian universities compiled by CENSIS, including, besides an overall score, specific ratings for the institution performances in the areas of services, scholarships, structures, communication and digital services, international outlook, employability \cite{66}; data are available for 74 Italian universities, except those on employability, that are not given for 16 private institutions.
\end{itemize}
All the ranking scores, both overall and sectorial, are rescaled in the interval $[0,1]$. As a benchmark to validate our procedure of ranking reinterpretation, we use indicators related to the wealth of regions. Specifically, we consider data on territorial GDP per capita at purchasing power parity (PPP), obtained from the OECD National Accounts Statistics database (set ``USD per head, constant prices, constant PPP, base year 2015'') \cite{79} in the international case, and the indicator ``Average per capita income (EUR)'' from the ISTAT database in the Italian case. For both benchmark indicators, data from the most recent year available are considered, with missing values integrated by entries from at most 5 years before. The GDP per capita at PPP of the regions HU10 (Central Hungary) and PL11 (Mazovia) has been derived by an average, weighted by region populations, of the values for HU11 (Budapest) and HU12 (Pest) in the first case, PL91 (Warsaw) and PL92 (Mazovian region) in the second case.

\subsection*{Complex network construction}

Here we illustrate the methods to construct the two classes of complex networks employed as analysis tools in our work, namely territorial networks, based on the similarity between the regions to which the main university seats belong, and educational offer networks, based on the presence, in different universities, of degrees in the same educational areas.

\subsubsection*{Territorial networks} 

We start by constructing subnational area networks, whose nodes are represented by the regions of the chosen territorial partition, namely the OECD TL2 regions (with few TL3 exceptions) and the Italian provinces (\textit{province}). In principle, each region can be connected to any other, with the weight of an edge being determined by the Pearson correlation between the vectors of territorial indicators pertaining to the two connected regions. However, we choose to not consider a given edge in case the null hypothesis of uncorrelated sets of indicators cannot be rejected at a significance level of 1\% (i.e., when $p>10^{-2}$, according to the test based on comparing the sample Pearson correlation with the exact distribution of the correlation values between two random vectors, independent and normally distributed). Therefore, territorial networks are not complete, and edges can have both positive and negative values. Starting from the subnational area networks, we construct the proper territorial university networks, in which nodes represent universities, and edges between them are related to the similarity between their regions.

\subsubsection*{Educational offer networks} 

In this case, we start with a network whose nodes coincide with universities. The weight of the edge connecting two given institutions u and v is determined by the Dice index
\begin{equation}
    DSC_{uv} = \frac{2\left| \Gamma_u \cap \Gamma_v \right| }{\left| \Gamma_u \right| + \left| \Gamma_v \right|}
\end{equation}
between the sets $\Gamma_u$ and $\Gamma_v$ of their educational areas appearing in THE rankings (for the OECD case) or active degree categories (for the Italian case), with $|\dots|$ denoting the cardinality of a set. Even in this case, though the network is in principle complete, we perform a selection of edges, to provide it with a non-trivial topology. Specifically, for each node in the complete network, we identify the largest weight of the edges connected to it, and collect all these largest weights in a set $Q$. Then, we remove from the complete network all the edges between universities $u$ and $v$ such that $DSC_{uv}<\min Q$. By construction, this procedure does not leave isolated nodes.

\subsection*{Assortativity}

In the case of a binary network, the definition of assortativity with respect to a continuous attribute with values $x_i$ for each node $i$ reads \cite{42}
\begin{equation}
    r = \frac{ \sum_{ij} \left( A_{ij} - \frac{k_i k_j}{2m} \right) x_i x_j }{ \sum_{ij} \left( k_i \delta_{ij} - \frac{k_i k_j}{2m} \right) x_i x_j } ,
\end{equation}
with $k_i$ the node degrees, $A_{ij}$ the adjacency matrix and $m$ the total number of edges in the network. The values of assortativity range from $-1$ (maximally antiassortative network) to $+1$ (maximally assortative). A network consisting of two equal and fully connected components, with no connection between them and characterized by two different values of an attribute, is maximally assortative. A network consisting of two subsets with the same cardinality, each characterized by a specific attribute value, with no internal connection but with each node only connected to nodes of the other subset, is maximally antiassortative. If $r=0$, there is no relevant linear correlation between the values $x_i$ and $x_j$ of the attributes of nodes $(i,j)$ connected by an edge (notice, however, that this does not exclude the existence of nonlinear correlations). 

The above definition can be straightforwardly generalized to weighted networks \cite{43}, 
\begin{equation}
    r_w = \frac{ \sum_{ij} \left( w_{ij} - \frac{s_i s_j}{W} \right) x_i x_j }{ \sum_{ij} \left( s_i \delta_{ij} - \frac{s_i s_j}{W} \right) x_i x_j }  ,
\end{equation}
where $w_{ij}$ is the weight of the edge $(i,j)$, $s_i=\sum_j w_{ij}$ is the strength of node $i$, and $W=\sum_{ij} w_{ij}$. The expression of $r_w$ remains meaningful only in the case of positive weights. Therefore, before evaluating assortativity, we associate to territorial networks, which can contain negative-weight edges, an auxiliary subnetwork where only positive-weight edges are retained. 

The definition of assortativity for a weighted network is formally equivalent to the weighted Pearson correlation between two vectors of length $2m$, with $m$ the number of edges, whose entries coincide respectively with the attributes $x_i$ and $x_j$ of the nodes at the ends of each edge $(i,j)$; the contribution of a given pair $(x_i,x_j)$ to the overall correlation is determined by the weight $w_{ij}$ of the corresponding edge. The interpretation of assortativity in terms of a weighted Pearson correlation allows to associate to $r_w$ the standard error
\begin{equation}
    S_w = \sqrt{ \frac{1-r_w^2}{2m-2} } ,
\end{equation}
that is evaluated, along with the related $p$-value, based on the Student $t$-distribution hypothesis \cite{80}. The assortativity and the associated standard errors and p-values are computed through an algorithm implemented in the ``weights'' R library \cite{81}.

\subsection*{Community detection}

Community detection is performed using the Spin Glass algorithm \cite{82,83}. While the resolution $\gamma$ is treated as a free parameter, varied in the interval $[0.8,1]$ with step width $0.05$, the other parameters of the algorithm are fixed to default values, outlined in detail in the Supplementary Section~4. For a fixed set of parameters, we perform $K=100$ runs of the algorithm, each one with a different seed of the pseudorandom number generator. The partition in communities is then chosen by majority voting. However, in order to evaluate the stability of the detected partition in communities, we do not just count the frequency of the majority partition. Instead, we introduce a new stability criterion, that takes into account the similarity between different partitions $\{p_j\}_{(j=1,..,K)}$, based on the average Normalized Mutual Information
\begin{equation}
    \left\langle NMI \right\rangle = \frac{2}{K(K-1)} \sum_{a=1}^{K-1} \sum_{b=a+1}^K NMI(p_a,p_b) ,
\end{equation}
where $NMI(p_a,p_b)$ is the Normalized Mutual Information between a given pair of partitions, and $K(K-1)/2$ is the number of distinct pairs. The majority partition over $K=100$ runs is approved under the condition $\langle NMI \rangle \geq 0.90$, related to the general stability of the community detection. Moreover, the majority partition must satisfy the following further requirements:
\begin{itemize}
    \item it must not be trivial (i.e., consisting of a single community, coinciding with the whole network); 
    \item it must not be too fragmented, containing communities whose cardinality is less than 5\% of the cardinality of the whole network.
\end{itemize}
If the results obtained for 100 runs, at different values of the resolution $\gamma$, satisfy the above conditions, we choose the output with larger $\langle NMI \rangle$, and the majority partition corresponding to this choice is identified as the result of community detection. This scheme is employed to determine the most stable partition of universities based on territorial features and educational offer. In the first case, territorial community detection is determined on the subnational area network, where nodes represent regions, and then the university network is divided according to a partition in which a node (now coinciding with an institution) inherits community membership of the pertaining region.

Considering the different kinds of networks and the hierarchical community detection process, we use the following nomenclature for communities:
\begin{itemize}
    \item ``O'' labels communities of an OECD network;
    \item ``I'' labels communities of an Italian network;
    \item ``T'' labels communities of a territorial network;
    \item ``E'' labels communities of an educational offer network;
    \item communities labelled with the same number derive from the same community found at the first hierarchical level;
    \item the lower-case letters after the number are used to distinguish communities found at the second hierarchical level.
\end{itemize}
For example, communities OE1a and OE1b belong to an OECD educational offer network, and are obtained at the second hierarchical level by subdividing the same community found at the first hierarchical level. On the other hand, communities OE2b and OE3b are obtained from two different first-level communities, as they are labelled by different numbers; the fact that they are labelled with the same lower-case letter is accidental. 

\subsection*{Gaussian Mixture models}

To investigate the multimodal nature of the principal components, we apply a procedure based on fitting PC1 and PC2 with a family of unidimensional Gaussian Mixture models, characterized by a number $n_c$ of components that varies from 1 to 10. At fixed $n_c$, the Gaussian character of each component is checked by the Shapiro-Wilk test \cite{84}, in which the null hypothesis of Gaussian distribution is not rejected if its $p$-value is larger than 5\%. Finally, we retain only the multimodal Gaussian models in which all the $n_c$ components pass the test. A further selection can be made on the accepted candidate models by choosing the one that minimizes the AIC or BIC score \cite{85}. The numerical findings are discussed in the Supplementary Section~2.4.



\section*{Acknowledgements}

Code development/testing and results were obtained on the IT resources hosted at ReCaS data center. ReCaS is a project financed by the italian MIUR (PONa3\_00052, Avviso 254/Ric.).

\section*{Author contributions statement}

Conceptualization: LB, AM, NA, RB. Methodology: LB, AM, NA, RB. Investigation: LB. Visualization: LB, ADL, AM, NA. Supervision: RB. Writing—original draft: LB. Writing—review \& editing: LB, AM, NA, VA, MB, ADL, AL, ST, RB.

\section*{Additional information}

\textbf{Competing interests}: Authors declare that they have no competing interests.

\noindent\textbf{Data and materials availability}: The data that support the findings of this study are either publicly available on databases cited in the bibliography, or reported in Supplementary Data files. Computer code is available from the corresponding author upon reasonable request.


\end{document}